\documentclass[prb,twocolumn,showpacs,preprintnumbers,amsmath,amssymb]{revtex4}
\usepackage{bm}
\usepackage[dvips]{graphicx}

\begin{document}

\title{Long-range dynamics of  magnetic impurities coupled to a two-dimensional
Heisenberg antiferromagnet}

\author{Andreas L\"uscher}
\email[E-mail:]{andreas.luescher@a3.epfl.ch}
\author{Oleg P. Sushkov}
\email[E-mail:]{sushkov@phys.unsw.edu.au}
\affiliation{School of Physics, University of New South Wales, Sydney 2052,
Australia}

\begin{abstract}
We consider a two-dimensional Heisenberg antiferromagnet on a square lattice
with weakly coupled impurities, i.e. additional spins interacting with the host
magnet by a small dimensionless coupling constant $g\ll 1$. Using linear
spin-wave theory, we find that the magnetization disturbance at distance $r$
from a single impurity behaves as $\delta S^z \sim g/r$ for $1\ll r \ll 1/g$
and as $\delta S^z \sim 1/(gr^3)$ for $r \gg 1/g$. Surprisingly the
magnetization disturbance is inversely proportional to the coupling constant!
The interaction between two impurities separated by a distance $r$ is $\delta
\epsilon \propto g^2/r$ for $1\ll r \ll 1/g$ and $\delta \epsilon \propto
1/r^3$ for $r \gg 1/g$. For large distances, the interaction is therefore universal and
independent of the coupling constant. We have also found that the frequency of Rabi
oscillations between two impurities is logarithmically enhanced compared to the
decay width $\omega_{Rabi} \propto g^2\ln(1/gr)$ at $1\ll r \ll 1/g$. This
leads to a logarithmic enhancement for NMR and EPR line broadening. All these
astonishing results are due to the gapless spectrum of magnetic excitations in
the quantum antiferromagnet.

\end{abstract}

\date{\today}

\maketitle

\section{Introduction}
The interplay of magnetic impurities with strongly correlated
electron systems has attracted considerable attention over the past
decade. The discovery of high-temperature superconductors stimulated
studies of impurities in two-dimensional Mott-insulators with
long-range antiferromagnetic order~\cite{manousakis,dagotto}, where
the copper-oxide parent compounds are driven to superconductivity by
doping with holes or electrons. At low doping, the dopants are
localized and it is therefore insightful to study the limit of
isolated static holes, which have been realized
experimentally~\cite{cheong,ting,corti,caretta,vajk} and extensively
studied theoretically~\cite{nagaosa1,bulut,poilblanc,krivenko,nagaosa2,sandvik,irkhin}.
 There is a considerable body of work devoted
to impurity bonds and added spins~\cite{igarashi,murayama,kotov,irkhin}, a
generalization of the static hole case, and an unexpected behavior of the
impurity magnetic susceptibility at low temperature has been revealed quite
 recently~\cite{sachdev,vojita,hoglund,vojita1,sushkov}. There is also a
separate and very interesting Kondo-like problem of an
impurity in a magnetic system close to an $O(3)$ quantum critical
 point~\cite{sachdev,vojita,sushkov1,troyer}.
 However, in this work, we consider
impurities (added spins) in a system with long-range
antiferromagnetic order. The very unusual behavior we find, closely
related to that observed for the magnetic susceptibility and other
quantities in the
 papers~\cite{nagaosa1,nagaosa2,chern,hoglund,vojita1,sushkov}, is due to
gapless Goldstone spin-wave excitations.

The rest of the paper is organized as follows. In section \ref{sec:model} we
formulate the model, derive the spin-wave vertices and introduce an effective
Hamiltonian describing the interaction between impurities, which is
 explicitly calculated in sections \ref{sec:interaction} an \ref{sec:rabi} as a
function of the distance between impurities. A similar problem has been
considered earlier in Ref.~\onlinecite{nagaosa1}, but only the case of small
separations has been addressed, here we focus on the long-range behavior.
 The magnetization disturbance in the host antiferromagnet induced by a single
impurity is then calculated in section \ref{sec:mag}, and finally section VI
presents our conclusions.

\section{Model\label{sec:model}}
\begin{figure}[h]
\includegraphics[width=0.3\textwidth,clip]{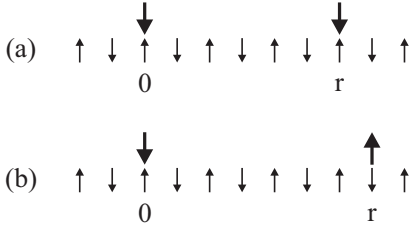}
\caption{Schematic picture of the antiferromagnetic host with two
antiferromagnetic impurities coupled to the origin and site {\bf r}
on the same sublattice (a) and on different sublattices (b).
\label{fig:lattice}}
\end{figure}

We consider one and two spin $\sigma=\frac{1}{2}$ impurities coupled
to an isotropic Heisenberg antiferromagnet ($J >0 $) on a square
lattice with lattice spacing $a=1$. The impurities are connected to
the origin and site $\bf r$ of the antiferromagnet. The Hamiltonian
of this system reads
\begin{equation} \label{eq:hamiltonian}
H=J \sum_{\left\langle i, j \right\rangle} {\bf S}_i \cdot {\bf S}_j
+ J' \left( {\bm \sigma}_1 \cdot {\bf S}_0 + {\bm \sigma}_2 \cdot
{\bf S}_r \right) \ .
\end{equation}
Here $\left\langle i, j \right\rangle$ denote nearest-neighbors, ${\bf S}_i$ is
a spin-$S$ operator at site ${\bf r}_i$ and ${\bf \sigma}_j$ are
spin-$\frac{1}{2}$ operators describing the impurities, which are either both
ferromagnetic ($J'<0$) or antiferromagnetic ($J'>0$).
 After integration over quantum fluctuations of the
system, one obtains the effective Hamiltonian for the interaction between
impurities
\begin{eqnarray} \label{eq:effham}
H_{eff}&=&J'[\langle S_0^z\rangle\sigma_1^z+\langle S_r^z\rangle\sigma_2^z]\nonumber\\
&+& \epsilon\left({\bf r}\right) \sigma_1^z \sigma_2^z+\left(M\left({\bf
r}\right) \sigma_1^+ \sigma_2^- + H.c.\right)\ .
\end{eqnarray}
The calculation of this effective Hamiltonian is the goal of the present work.
The diagonal interaction term $\epsilon\left({\bf r}\right)$, as well as the
off-diagonal term $M\left({\bf r}\right)$ are different, depending on whether
the impurities are coupled to the same or different sublattices, see
Fig.~\ref{fig:lattice}(a) and (b) respectively. Without loss of generality, we
place the origin on the up sublattice (sublattice ``a'') and indicate the
sublattice the second impurity is coupled to by the corresponding letter, see
Figs.~\ref{fig:lattice}. In this way, ``a'' (``b'') refers to impurities
coupled to the same (different) sublattice.
We study the weak coupling regime $\left|J'\right| \ll J$, so that the
dimensionless coupling constant is small
\begin{equation}
\label{gc} g=\left|J'\right|/(2\sqrt{2}J) \ll 1 \ .
\end{equation}

To account for quantum fluctuations, we treat the host antiferromagnet in the
linear spin-wave approximation where excitations are described by operators
$\alpha_{\bf q}^\dag$ and $\beta_{\bf q}^\dag$ creating spin-waves with
$S^z=-1$ and $S^z=+1$ respectively, see Ref.~\onlinecite{manousakis} for a
review. In this approximation, the Hamiltonian (\ref{eq:hamiltonian}) can be
decomposed as $H=H_0 + V$, where
\begin{eqnarray}
&H_0&= E_0 + 4JS \sum_{\bf q} \omega_{\bf q} \left( \alpha_{\bf
q}^\dag \alpha_{\bf q} + \beta_{\bf q}^\dag \beta_{\bf q} \right) +
J' S \left(\sigma_1^z  \pm \sigma_2^z \right), \nonumber \\
\label{hamV} &V&=H_1^{(a)} + H_2^{(a),(b)}\ , \text{ with}
\\[0.2cm]
\label{eq:hamup}
&H_i^{(a)}&=\nonumber\\
&-&J' \sigma_i^z \frac{2}{N}\sum_{{\bf p}, {\bf q}} e^{i\left({\bf
p}-{\bf q}\right) \cdot \bf{r_i}} \left( u_{\bf p} \alpha_{\bf
p}^\dag + v_{\bf p}\beta_{\bf p} \right) \left( u_{\bf q}
\alpha_{\bf q}
+ v_{\bf q} \beta_{\bf q}^\dag\right)\nonumber\\
&+& J' \sqrt{\frac{S}{N}} \left( \sigma_i^+ \sum_{\bf q} e^{i{\bf q}
\cdot \bf{r_i}} \left( u_{\bf q} \alpha_{\bf q}^\dag + v_{\bf
q}\beta_{\bf q}\right) +
\text{h.c.} \right)\ , \\
\label{eq:hamdown}
&H_i^{(b)}& =\nonumber\\
&& J' \sigma_i^z \frac{2}{N}\sum_{{\bf p}, {\bf q}} e^{i\left({\bf
p}- {\bf q}\right) \cdot \bf{r_i}} \left( u_{\bf p} \beta_{\bf
p}^\dag + v_{\bf p} \alpha_{\bf p} \right) \left( u_{\bf q}
\beta_{\bf q} + v_{\bf q} \alpha_{\bf q}^\dag
\right)\nonumber\\
&+& J' \sqrt{\frac{S}{N}} \left( \sigma_i^+ \sum_{\bf q} e^{i {\bf
q} \cdot \bf{r_i}} \left( u_{\bf q} \beta_{\bf q} + v_{\bf q}
\alpha_{\bf q}^\dag \right) + \text{h.c.}  \right) \ .
\end{eqnarray}
Here $E_0$ is the ground state energy of the antiferromagnetic host
and $N$ the number of sites. The upper (lower) sign in the $\pm$
expression refers to the situation where the second impurity is
coupled to a spin on sublattice ``a'' (sublattice ``b'').  This
convention is used throughout the whole paper. The Bogoliubov
parameters $u_{\bf q}$ and $v_{\bf q}$ are given by
\begin{align*}
u_{\bf q} &= \sqrt{\frac{1}{2\omega_{\bf q}}+\frac{1}{2}} & v_{\bf
q} &= -\text{sgn} \left(\gamma_{\bf q}\right)
\sqrt{\frac{1}{2\omega_{\bf q}}-\frac{1}{2}}\ ,
\end{align*}
with $\omega_{\bf q} = \sqrt{1-\gamma_{\bf q}^2}$ and $\gamma_{\bf
q} = \frac{1}{2}\left(\cos{q_x}+\cos{q_y}\right)$, see
Ref.~\onlinecite{manousakis}. In this notation, the spin-wave
dispersion is $\epsilon_{\bf q}=4JS\omega_{\bf q}$. The interaction
Hamiltonians (\ref{eq:hamup}) and (\ref{eq:hamdown}) generate one
and two spin-wave vertices summarized in Table~\ref{tab:vertices}.

\begin{table} \centering \caption{\label{tab:vertices}
\emph{Spin-wave vertices for an impurity coupled to site {\bf r} on sublattice
``a'' or ``b''.}}\begin{ruledtabular}
\begin{tabular}{cccc}
 Symbol & Operator & Factor for ${\bf r}$ on ``a'' & or ${\bf
r}$ on ``b''\\ \hline \\
$\blacktriangle$ & $\sigma^z\alpha_{\bf p}^\dag\alpha_{\bf q}$ &
$-\tfrac{2J'}{N} u_{\bf p} u_{\bf q}e^{i \left({\bf p}- {\bf q}\right) \cdot
{\bf r}} $ &
  $\tfrac{2J'}{N} v_{\bf p} v_{\bf q}e^{-i \left({\bf p}-  {\bf q} \right) \cdot {\bf r}}$ \\
$\vartriangle$ & $\sigma^z\beta_{\bf p}^\dag
\beta_{\bf q}$ & $-\tfrac{2J'}{N} v_{\bf p} v_{\bf q}e^{-i \left({\bf p}- {\bf
q}\right)\cdot {\bf r}} $ &
  $\tfrac{2J'}{N} u_{\bf p} u_{\bf q} e^{i \left({\bf p}- {\bf q} \right) \cdot {\bf r}}$ \\
$\blacklozenge$ & $\sigma^z\alpha_{\bf p}^\dag
\beta_{\bf q}^\dag$ & $-\tfrac{2J'}{N} u_{\bf p} v_{\bf q}e^{i \left({\bf p}- {\bf
q} \right) \cdot {\bf r}}$ &
  $\tfrac{2J'}{N} v_{\bf p} u_{\bf q}e^{-i \left({\bf p}- {\bf q} \right) \cdot {\bf r}}$ \\
$\lozenge$ & $\sigma^z\beta_{\bf p} \alpha_{\bf
q}$ & $-\tfrac{2J'}{N} v_{\bf p} u_{\bf q} e^{i \left({\bf p}- {\bf q} \right)
\cdot {\bf r}}$ &
  $ \tfrac{2J'}{N} u_{\bf p} v_{\bf q} e^{-i \left({\bf p}- {\bf q} \right) \cdot {\bf r}}$ \\
\\$\bullet$ & $\sigma^+ \alpha_{\bf q}^\dag$
& $J'\sqrt{\tfrac{S}{N}} u_{\bf q}e^{i {\bf q} \cdot {\bf r}} $
 & $ J'\sqrt{\tfrac{S}{N}} v_{\bf q}e^{i {\bf q} \cdot {\bf r}} $ \\
$\circ$ & $\sigma^-\alpha_{\bf q}$ & $
J'\sqrt{\tfrac{S}{N}} u_{\bf q} e^{-i {\bf q} \cdot {\bf r}}$ &
  $J'\sqrt{\tfrac{S}{N}} v_{\bf q} e^{-i{\bf q} \cdot {\bf r}} $ \\
$\blacksquare$ & $\sigma^-\beta_{\bf
q}^\dag$ &$J'\sqrt{\tfrac{S}{N}} v_{\bf q} e^{-i {\bf q} \cdot {\bf r}}$ &
  $ J'\sqrt{\tfrac{S}{N}} u_{\bf q} e^{-i {\bf q} \cdot {\bf r}} $ \\
$\square$ & $\sigma^+ \beta_{\bf q}$ &$
J'\sqrt{\tfrac{S}{N}} v_{\bf q} e^{i {\bf q} \cdot {\bf r}}$ &
  $J'\sqrt{\tfrac{S}{N}} u_{\bf q} e^{i{\bf q} \cdot {\bf r}} $ \\
\end{tabular}\end{ruledtabular} \end{table}

Let us return to the effective Hamiltonian (\ref{eq:effham}). The first two
terms are obvious and do not require calculations. They simply generate three
energy levels, $E_0=-J'\langle S^z\rangle$, $E_1=E_{\overline{1}}=0$, and
$E_2=J'\langle S^z\rangle$. To be specific, let us consider the case shown in
Fig.~\ref{fig:lattice}(a), with impurities coupled to the same sublattice, then
\begin{eqnarray}
\label{012}
|0\rangle&=& |\downarrow,\downarrow\rangle \ ,\nonumber\\
|1\rangle&=& |\downarrow,\uparrow\rangle \ ,\nonumber\\
|\overline{1}\rangle&=& |\uparrow,\downarrow\rangle \ ,\nonumber\\
|2\rangle&=& |\uparrow,\uparrow\rangle \ .
\end{eqnarray}
Only the ground state $|0\rangle$ is the true stationary quantum state. The
states $|1\rangle$, $|{\overline{1}}\rangle$, and $|2\rangle$ decay to the
ground state with emission of spin-waves. Using the Fermi golden rule and the
decay matrix elements presented in Tab.~\ref{tab:vertices}, one finds the
following widths of excited states with respect to the emission of magnons
\begin{eqnarray}
\label{gsw}
&&\Gamma_1= 2g^2 J \ ,\nonumber\\
&&\Gamma_2= 2\Gamma_1\ .
\end{eqnarray}
The three-level system is well defined, since $E_1-E_0 \gg \Gamma_1$. But the
diagonal and off-diagonal interaction energies $\epsilon\left({\bf r}\right)$
and $M\left({\bf r}\right)$ in the effective Hamiltonian (\ref{eq:effham}) have
limited meaning because of the finite lifetime. Only in the ground state
$|0\rangle$, the diagonal interaction energy is well defined. In the spin flip
states $|1\rangle$, $|{\overline{1}}\rangle$,
 and $|2\rangle$
the diagonal interaction energy does not make much sense, because we will see
that it is always much smaller than the corresponding decay width. However,
there is a regime where the off-diagonal interaction $M\left({\bf r}\right)$ is
larger than the decay width and hence leads to Rabi oscillations between states
$|1\rangle$ and $|{\overline{1}}\rangle$.

\section{Diagonal interaction {\large $\epsilon$}$\left({\bf r}\right)$
between impurities\label{sec:interaction}} In the leading order of the
$\frac{1}{S}$-expansion, the interaction energy $\epsilon\left({\bf r}\right)$
arises in second, third and fourth orders of perturbation theory, describing
the exchange of two spin-waves between impurities. The corresponding
contributions in usual Rayleigh-Schr\"odinger perturbation theory
are~\cite{Land}
\begin{eqnarray}
\label{234} \delta \epsilon_2&=&\sum_{n \ne 0}\frac{\langle 0|V|n\rangle\langle
n|V|0\rangle}{(\epsilon_0-\epsilon_n)} \ ,\nonumber\\
\delta \epsilon_3&=&\sum_{n,m \ne 0}\frac{\langle 0|V|n\rangle\langle
    n|V|m\rangle\langle m|V|0\rangle}
{(\epsilon_0-\epsilon_n)(\epsilon_0-\epsilon_m)} \ , \\
\delta \epsilon_4&=&\sum_{n,m,k \ne 0}\frac{\langle 0|V|n\rangle\langle
    n|V|m\rangle\langle m|V|k\rangle\langle k|V|0\rangle}
{(\epsilon_0-\epsilon_n)(\epsilon_0-\epsilon_m)(\epsilon_0-\epsilon_k)}
\ , \nonumber
\end{eqnarray}
where $V$ is the perturbation~(\ref{hamV}). In general, the
expressions for third and fourth order energy corrections are more
complex than those in~(\ref{234}). The complication is due to
contributions similar to~(\ref{234}), but in which intermediate
states $m,n,k$ coincide with the initial state $|0\rangle$.
Fortunately, the perturbation~(\ref{hamV}) does not allow such
intermediate states and therefore Eqs.~(\ref{234}) are valid.
\begin{figure}
\includegraphics[width=0.25\textwidth,clip]{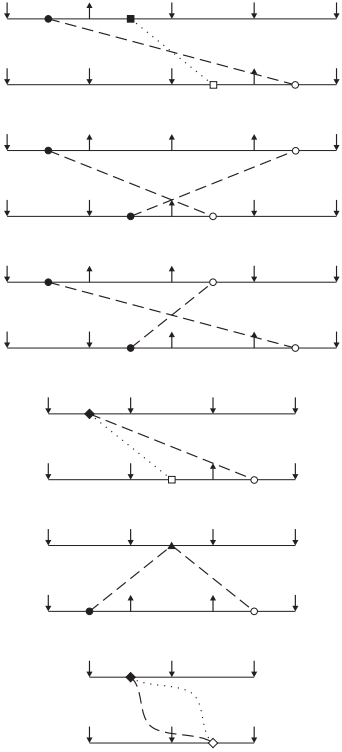}
\caption{\emph{Diagrams describing the $\frac{1}{S}$-corrections for
antiferromagnetic impurities coupled to sites on the same sublattice
(case ``a'').
 The spin-wave vertices are summarized in
Tab.~\ref{tab:vertices}, dashed and dotted lines represent $\alpha$- and
$\beta$-spin-waves respectively.} \label{fig:diagramssame}}
 \end{figure}
The matrix elements $\langle n|V|m\rangle$ are given in
Table~\ref{tab:vertices} and it is convenient to represent the
corrections (\ref{234}) by diagrams where each vertex corresponds to
some particular matrix element $\langle n|V|m\rangle$.
\begin{figure}
\includegraphics[width=0.25\textwidth,clip]{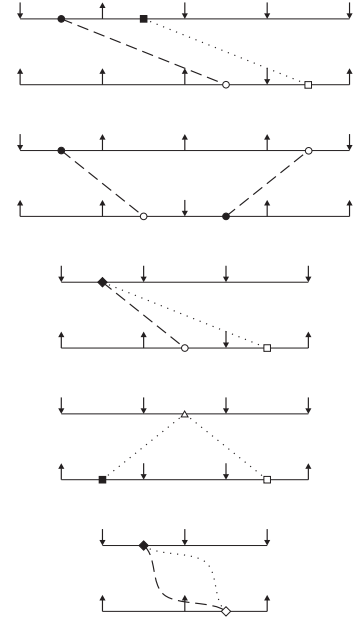}
\caption{\emph{Diagrams describing the $\frac{1}{S}$-corrections for
antiferromagnetic impurities coupled to sites on different sublattices (case
``b''). The notation is the same as in Fig.~\ref{fig:diagramssame}.}
\label{fig:diagramsdifferent}}
\end{figure}
The diagrams describing the $\frac{1}{S}$-corrections are shown in Figs.~\ref{fig:diagramssame}
and \ref{fig:diagramsdifferent} for antiferromagnetic impurities ($J' > 0$) coupled to the same
 and different sublattices  respectively.
 Ferromagnetic impurities ($J' < 0$) generate similar
diagrams with interchanged $\alpha$- and $\beta$-spin-waves.
The first diagram in
Fig.~\ref{fig:diagramssame} for instance represents the expression
\begin{equation*}
\frac{ -2 {J'}^4 S^2 u_{\bf p}^2 v_{\bf q}^2 e^{-i \left({\bf p}-
{\bf q}\right) \cdot {\bf r}}}{N^2 \left(\left|J'\right|S+4JS
\omega_{\bf p}\right) 4J S \left(\omega_{\bf p}+ \omega_{\bf
q}\right) \left(\left|J\right|'S+4JS\omega_{\bf p}\right)}\ ,
\end{equation*}
with a factor $2$ taking into account the similar process with exchanged impurities.

 Summing these
contributions, we obtain the interaction energy in the leading $\frac{1}{S}$-order
\begin{equation}\label{eq:int}
\delta \epsilon \left({\bf r}\right) = \frac{4J{J'}^2}{SN^2} \sum_{{\bf p}, {\bf q}}
\frac{A_{{\bf p}, {\bf q}} e^{i\left({\bf p}-{\bf q}\right) \cdot {\bf r}}}{B_{{\bf p}, {\bf
q}}}  \ ,
\end{equation}
with
\begin{eqnarray*}
A_{{\bf p}, {\bf q}}^{(a)} &=& {J'}^2 u_{\bf q}^2 \left(u_{\bf p}^2 \left(\omega_{\bf p} +
\omega_{\bf q}\right)^2- 2 v_{\bf p}^2 \omega_{\bf q}^2\right) \\ &&\quad + 8J{J'} u_{\bf q}^2
\omega_{\bf p} \omega_{\bf q} \left(u_{\bf p}^2\left(\omega_{\bf p}+
\omega_{\bf q}\right) -2 v_{\bf p}^2 \omega_{\bf q}\right)  \\
&&\quad\quad - 32 J^2 u_{\bf p}^2
v_{\bf q}^2 \omega_{\bf p}^2 \omega_{\bf q}^2  \\
A_{{\bf p}, {\bf q}}^{(b)} &=& -2 u_{\bf p} u_{\bf q} v_{\bf p}
v_{\bf q} \omega_{\bf p} \omega_{\bf q} \\
B_{{\bf p}, {\bf q}}^{(a)} &=& \left(\omega_{\bf p}+\omega_{\bf q}\right)
\left(\left|J'\right|+4J \omega_{\bf p}\right)^2
\left(\left|J'\right|+4J \omega_{\bf q}\right)^2 \\
B_{{\bf p}, {\bf q}}^{(b)} &=& \left(\omega_{\bf p}+\omega_{\bf q}\right)
\left(\left|J'\right|+4J \omega_{\bf p}\right) \left(\left|J'\right|+4J \omega_{\bf q}\right)\ .
\end{eqnarray*}
The sum extends over momenta in the magnetic Brillouin zone. In the case of ferromagnetic
impurities ($J'<0$), the Bogoliubov parameters in (\ref{eq:int}) have to be interchanged.

For large distances, $r \gg 1$, the interaction energy comes from
small momenta, $p,q \approx 1/r \ll 1$. One can therefore
approximate the Bogoliubov parameters and the dispersion by
\begin{align}\label{eq:approximation}
\omega_{\bf q} \approx \frac{q}{\sqrt{2}}\ , && u_{\bf q} \approx
\sqrt{\frac{1}{\sqrt{2} q}} && \text{and} && v_{\bf q} \approx
-\sqrt{\frac{1}{\sqrt{2} q}} \ ,
\end{align}
and hence simplify the interactions (\ref{eq:int})
\begin{eqnarray}
\label{e12} \delta \epsilon^{(a)} \left({\bf r} \right) &\approx&  \frac{J
g^2}{S}\frac{2\sqrt{2}}{\left(2 \pi\right)^4} \int \frac{\left(g^2 - q p\right)
e^{i \left({\bf p}- {\bf q} \right) \cdot {\bf r}}d^2p\;d^2q } {\left(p+q\right)
\left(g+ p\right)^2 \left(g+q\right)^2} \ , \nonumber \\
\delta \epsilon^{(b)} \left({\bf r} \right) &\approx& - \frac{J g^2}{S}
\frac{2\sqrt{2}}{\left(2 \pi\right)^4} \int \frac{e^{i \left({\bf p}- {\bf q}
\right) \cdot {\bf r}} d^2p\;d^2q} {\left(p+q\right) \left(g+ p\right)
\left(g+q\right)}\ .
\end{eqnarray}
Here $g$ is the dimensionless coupling constant (\ref{gc}). Since
both  integrals are ultraviolet convergent, we extend the
integration domain to infinity. There is no difference between
ferro- and antiferromagnetic  ($J'=\pm |J'|$) impurities in this
approximation.

We first consider \emph{very large} distances $r\gg 1/g$. In this case, momenta
in (\ref{e12}) are limited by $p,q \ll g$ and the corrections to the ground
state energy are equal to
\begin{equation*} \delta \epsilon \left({\bf r} \right)
\approx \pm \frac{J}{S} \frac{2\sqrt{2}}{\left(2 \pi\right)^4} \iint
\frac{e^{i \left({\bf p}- {\bf q} \right) \cdot {\bf r}} }
{\left(p+q\right)}d^2p\;d^2q\ ,
\end{equation*}
where according to our convention the plus sign corresponds
 to $\delta \epsilon^{(a)} \left({\bf r}\right)$ (same sublattice)
and the minus sign corresponds to $\delta \epsilon^{(b)} \left({\bf r} \right)$
(different sublattices). A first integration over angles gives two Bessel
functions $J_0$, which can be integrated using $\int_0^\infty \frac{J_0\left(p
r \right)
 J_0\left(q r \right) p q dp dq}{p + q}
 = \frac{1}{r^3} \frac{\pi}{16}$, yielding
\begin{equation}
\label{dev} \delta \epsilon \left({\bf r}\right) = \pm \frac{1}{16 \sqrt{2}
\pi} \frac{J}{Sr^3}\quad\left(r\gg 1/g\right)\ .
\end{equation}
Interestingly, in this limit, the interaction between impurities
is independent of their coupling to the antiferromagnetic host $J'$.

In the case of \emph{intermediate} distances between impurities,
$1 \ll r \ll 1/g$, a similar calculation using $\int_0^\infty \frac{J_0\left(p r
\right) J_0\left(q r \right) dp dq}{p + q}
 = \frac{1}{r} \frac{\pi}{2}$, leads to
\begin{equation}
\label{dei}
 \delta \epsilon \left({\bf r} \right)
 = -\frac{g^2}{2\sqrt{2}\pi} \frac{J}{Sr} \quad
\left(1 \leq r \ll 1/g\right)\ .
 \end{equation}
If the separation between the impurities is small, $r \approx 1$, the
approximation (\ref{eq:approximation}) for the Bogoliubov parameters and the
dispersion is no longer valid and the integrals in (\ref{eq:int}) have to be
calculated numerically. In order to circumvent finite-size effects, we
extrapolate the results obtained from lattices with up to $N^2=100 \times 100$
sites by a polynomial $\delta\epsilon\left(r\right)=a(r)+b(r)/N+c(r)/N^2$. Even
for very small distances, the interaction between additional spins is
surprisingly well fitted by the intermediate distance attractive behavior
(\ref{dei}). The situation is analogous to the magnetization disturbance, see
section V, for which Fig.~\ref{fig:mag} provides an illustration of
qualitatively similar fits. The attractive interaction (\ref{dei}) found in the
weak coupling regime is independent of the sublattice. In comparison, in the
strong coupling limit $\left|J'\right| \ge J$ an attractive interaction is
found for nearest neighbors, but the interaction between
 impurities on next-nearest sites is repulsive~\cite{bulut,nagaosa1}.

In the above calculations, we assumed that the impurities are in the ground
state, see Eq.~(\ref{012}). To check the kinematic structure of the diagonal
interaction in the effective Hamiltonian (\ref{eq:effham}) one has to perform
similar calculations for states $|1\rangle$, $|{\overline{1}}\rangle$, and
$|2\rangle$, see (\ref{012}). In the case where both impurities are flipped
with respect to their ground state configuration, the interaction is the same
as in the ground state, since the situation corresponds to ferromagnetic
coupling with $J'>0$. For an excited state with only one flipped impurity, a
calculation analogous to (\ref{eq:int}) shows that the interaction has opposite
sign and hence justifies the kinematic structure of the $\sigma_1^z \sigma_2^z$
term in (\ref{eq:effham}). We emphasize (see also end of section
\ref{sec:model}), that because of the finite lifetime, the diagonal interaction
in the spin-flipped states has limited meaning.

\section{Off-diagonal interaction $M\left({\bf r}\right)$
and Rabi oscillations between impurities\label{sec:rabi}} 
\begin{figure}
\includegraphics[width=0.25\textwidth,clip]{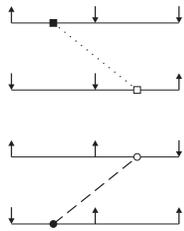}
\caption{\emph{Diagrams describing Rabi oscillations:
 The system is brought into a state where one spin is
flipped, for example during an NMR experiment and oscillates between excited
states by exchanging spin-waves.} \label{fig:diagramssrabi}} \end{figure}

If a magnetic
impurity is flipped with respect to the ground state configuration, e.g. in a
nuclear magnetic resonance (NMR) or electron paramagnetic resonance (EPR)
experiment, then the excited state has a finite lifetime due to emission of
magnons, see first of Eqs. (\ref{gsw}). However, there is another mechanism of
spin relaxation, due to presence of distant similar magnetic impurities: a
spin-wave exchange leads to Rabi oscillations between two impurities. In the
effective Hamiltonian (\ref{eq:effham}) this process is described by the
off-diagonal term $M\left({\bf r}\right)$. Diagrams for $M\left({\bf r}\right)$
are shown in Fig.~\ref{fig:diagramssrabi}. Rabi oscillations can only be
observed between impurities coupled to spins on the \emph{same} sublattice, so
\begin{eqnarray}
&&M^{(a)}\left({\bf r}\right)\ne 0 \ , \nonumber\\
&&M^{(b)}\left({\bf r}\right)=0 \ . \nonumber
\end{eqnarray}
The oscillation frequency of this two-level system is proportional to the real part of the mixing
matrix element $M=\left\langle \uparrow,\downarrow\right|H\left|\downarrow,\uparrow\right\rangle$,
since the probability $P\left(t\right)$ to find the system in a state
 with flipped impurities after the time $t$ is equal to
$P\left(t\right)=\sin^2 \left( \text{Re }M\; t\right)$. Using Tab.~\ref{tab:vertices} to evaluate
the diagrams
 shown in Fig.~\ref{fig:diagramssrabi}, we find
\begin{equation*}
M^{(a)}\left(\bf r\right) =  \frac{{J'}^2}{N} \sum_{\bf q} e^{i {\bf q} \cdot {\bf r}} \left(
\frac{v_{\bf q}^2}{J'-4J \omega_{\bf q}+i\delta} - \frac{u_{\bf q}^2}{J'+4J\omega_{\bf q}}
\right).
\end{equation*}
For large distances $r \gg 1$ we use the approximate Bogoliubov parameters
(\ref{eq:approximation}). The mixing element becomes
\begin{eqnarray*}
M^{(a)}\left(\bf r\right) &=& -J \frac{g^2}{\pi} \int J_0\left(qr\right) \left( \frac{2q}{q^2-g^2}
 +i \pi \delta \left(q-g\right)\right) dq \\
 &=& J g^2 \left( Y_0\left(gr\right)-i J_0\left(gr\right)\right)\ ,
\end{eqnarray*}
where $Y_0$ is the Neumann function. For \emph{very large} distances between
impurities, $r \gg 1/g$, the real part of the mixing matrix element is
comparable to the imaginary one and both are much smaller than the width
(\ref{gsw}). In this case, there are no Rabi oscillations. However, in the
\emph{intermediate} regime, $1 \ll r \ll 1/g$, the real part
\begin{equation}
\label{rem}
\text{Re } M^{(a)}\left(\bf r\right) = J g^2 \frac{2}{\pi} \ln{gr}
\end{equation}
is logarithmically enhanced compared to the imaginary part and compared to the spin-wave width
(\ref{gsw}). Thus, in this regime, Rabi oscillations between impurities are well pronounced and
this mechanism gives the main contribution to the effective width of the magnetic resonance line
\begin{equation}\label{geff}
\Gamma_{eff}\approx J\frac{4g^2}{\pi^2}\left| \ln{gr}\right| \ .
\end{equation}

\section{Magnetization cloud around an impurity\label{sec:mag}}
\begin{figure} 
\includegraphics[width=0.45\textwidth,clip]{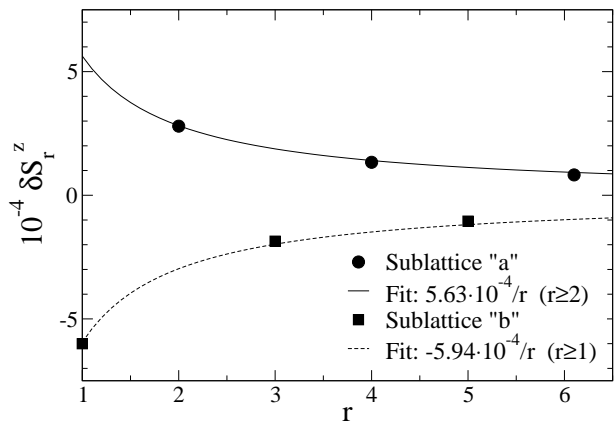}
\caption{\emph{The points show the results of a numerical calculation of the
variation of the host magnetization ($S=\frac{1}{2}$) at site $r$ induced by a
ferromagnetic impurity coupled to the origin by $J'=-0.01$. The curves show
fits of these points according to $\delta S_r^z=F/r$, where $F$ is the fitting
coefficient. Fitted values of $F$ (see inset) are very close to the value
$F=5.62\cdot10^{-4}$ that follows from Eq. (\ref{dsi}). \label{fig:mag}}}
\end{figure}

An interesting question is how an additional spin influences the magnetic order in the host
antiferromagnet, described by the staggered magnetization.  Let us consider an impurity $\sigma_1$
at the origin and a local magnetic field $h$ on site $r$. The Hamiltonian reads
\begin{equation}
\label{hi}
H = J \sum_{\left\langle i, j\right\rangle} {\bf S}_i \cdot {\bf S}_j + J' {\bm \sigma}_1 \cdot
{\bf S}_0 + h S_r^z \ ,
\end{equation}
and the variation of the expectation value of the local magnetization $\delta S_r^z$ in the ground
state is given by
\begin{equation}
\label{sh}
\delta S_r^z = \frac{\partial \delta E}{\partial h}\ ,
\end{equation}
where $\delta E$ is the part of the energy dependent on $J'$. Now we can
consider the same framework as for the calculation of the interaction between
impurities. Using Eq. (\ref{sh}) together with the Rayleigh-Schr\"odinger
perturbation theory, one can see that
\begin{equation*}
\delta S_r^{z} = \mp 2\delta \epsilon \left({\bf r}\right)\ ,
\end{equation*}
where $\delta \epsilon$ is the energy correction described by those diagrams in
Figs.~\ref{fig:diagramssame} and \ref{fig:diagramsdifferent} without spin flip of the second
impurity. These diagrams give the following explicit expressions for the magnetization variation
\begin{equation} \label{eq:mag}
\delta S_r^{z} = \frac{2J'}{S N^2} \sum_{{\bf p}, {\bf q}} \frac{ C_{{\bf p}, {\bf q}}
e^{i\left({\bf p}-{\bf q}\right) \cdot {\bf r}}}{D_{{\bf p}, {\bf q}}} \ ,
\end{equation}
 with
\begin{eqnarray*}
C_{{\bf p}, {\bf q}}^{(a)} &=& 2 J' u_{\bf q}^2 v_{\bf p}^2 \omega_{\bf q} - J' u_{\bf p}^2
u_{\bf q}^2 \left(\omega_{\bf p}+\omega_{\bf p}\right)+ 8 J u_{\bf q}^2 v_{\bf
p}^2 \omega_{\bf p} \omega_{\bf q}\\
C_{{\bf p}, {\bf q}}^{(b)} &=& -8 J u_{\bf p} u_{\bf q} v_{\bf
p} v_{\bf q} \omega_{\bf p} \omega_{\bf q}\\
D_{{\bf p}, {\bf q}}&=& \left(\omega_{\bf p}+\omega_{\bf q}\right) \left(\left|J'\right|+4J
\omega_{\bf p}\right) \left(\left|J'\right|+4J \omega_{\bf q}\right)
\end{eqnarray*}assuming $J' > 0$. The expressions for the ferromagnetic case ($J' <0$) are obtained by a change
of the global sign and interchanged Bogoliubov parameters in (\ref{eq:mag}). Using the
approximations~(\ref{eq:approximation})
 for $r\gg1$, we find that the magnetization disturbance is equal to
\begin{equation}
\label{dsv} \delta S_r^z = \pm \frac{1}{8 \sqrt{2} \pi}\frac{J}{\left|J'\right|}
\frac{1}{Sr^3}=\pm \frac{1}{32 \pi g} \frac{1}{Sr^3} \quad\left(r \gg
1/g\right)\ ,\end{equation}
in the \emph{very large} distance limit, and
\begin{equation}
\label{dsi} \delta S_r^z =
 \pm \frac{1}{8\sqrt{2}\pi}\frac{\left|J'\right|}{J}  \frac{1}{Sr}
= \pm \frac{g}{4 \pi} \frac{1}{Sr} \quad\left(1 \ll r \ll
1/g\right)\ .
\end{equation}
for the \emph{intermediate} region. In the case of small separation from the
impurity, one has to calculate the integrals in (\ref{eq:mag}) numerically.
Using the same finite-size extrapolation scheme as in section
\ref{sec:interaction} we find that the variation of the magnetization in the
vicinity of the impurity is described by Eq. (\ref{dsi}) down to $r=1$.
Fig.~\ref{fig:mag} displays the variation of the magnetization calculated
numerically at $J'=-0.01$. The results are very close to the analytical
expression (\ref{dsi}).
In agreement with Ref.~\onlinecite{nagaosa1} an added spin always enhances the
N\'eel order in the host magnet, independent of the sign of the exchange
coupling. In contrast to a vacancy, this enhancement is not limited to nearest
neighbor sites~\cite{bulut,sandvik}, but extends over the whole magnet. It is
also interesting to compare our result for an added spin to an
in-plane impurity considered in Ref.~\onlinecite{irkhin}. If the impurity
is placed inside the host, it weakens the N\'eel order of the surrounding spins,
but the magnetization disturbance also decreases as $1/r^3$ for $r\gg1$~\cite{irkhin}.

\section{Conclusion\label{sec:conclusion}}
To conclude, we have studied the long-range dynamics of one and
two spin-$\frac{1}{2}$
 impurities in a two-dimensional Heisenberg antiferromagnet with on site
 spin-$S$ treated in the linear spin-wave approximation.
  The impurities are assumed to be weakly
coupled to the host magnet by a small dimensionless coupling constant $g$.
 A systematical
treatment of the corrections contributing
to the leading order of the $\frac{1}{S}$-expansion
leads to non-trivial long-range dynamics.
The interaction between two impurities
can be separated into two regimes: For very large separations ($r \gg 1/g$) it is universal
(independent of $J'$) and decreases as $1/r^3$. The interaction is repulsive
 (attractive) for impurities coupled to the same (different) sublattices.
In an intermediate region $1 \ll r \ll 1/g$, the interaction decreases only as
$1/r$ and is attractive, independent of the sign of the exchange couplings. It is
shown that Rabi oscillations between impurities
coupled to spins on the same sublattice are possible and well pronounced in the intermediate
regime. The effective Hamiltonian decsribing the interaction in terms of the impurity spins
is derived. It exhibits an xyz anisotropy which leads to NMR and EPR line broadening.
The magnetization disturbance in the host magnet induced by a single impurity
is analyzed in the same framework. It is shown that the disturbance exhibits
behaviors similar to the interaction energy, always enhancing the magnetic
order in the antiferromagnetic host. Numerical results indicate that the
intermediate regimes can be extended down to $r\approx1$.

\end{document}